# *Establishment of the fundamental phase-to-polarization link in classical optics*

Xiaoyu Weng[1]*, Xiumin Gao[2]*, Guorong Sui[2], Qiang Song[2], Xiangmei Dong[2], Junle Qu[1,3] and Songlin Zhuang[2]

*1 College of Physics and Optoelectronic Engineering, Shenzhen University, Shenzhen, 518060, China.*

*2 Engineering Research Center of Optical Instrument and System, Ministry of Education, Shanghai Key Lab of Modern Optical System, School of Optical-Electrical and Computer Engineering, University of Shanghai for Science and Technology, 516 Jungong Road, Shanghai 200093, China.*

*3 Key Laboratory of Optoelectronic Devices and Systems of Ministry of Education and Guangdong Province, College of Physics and Optoelectronic Engineering, Shenzhen University, Shenzhen, 518060, China.*

\* Correspondence and requests for materials should be addressed to X.W. (email: xiaoyu@szu.edu.cn )
or to X.G. (email: gxm@usst.edu.cn ).

# Abstract:

Vector polarization induced by a change in scalar phase has been far beyond our understanding of the relationship between polarization and phase in classical optics due to the entanglement of inherent polarized modes of light beams. To overcome this limitation, we establish this miraculous relationship by the principle of phase vectorization, which can transform three particular phases into linear, circular and elliptical polarizations. A polarized-spatial light modulator based on the principle of phase vectorization can be achieved using a phase-only spatial light modulator, which not only enables pixel-level polarization manipulation of light beams in a real-time dynamic way but also simultaneously retains complete phase control. This work demonstrates the phase-to-polarization link and reports the creation of polarized-spatial light modulator, which will transform our view of the natural properties of light beam and pave the way for the era of vector optics.

# Introduction

Amplitude, phase and polarization characterize a light beam in classical optics. The wavefront information of the light beam is carried by amplitude and phase, and polarization refers to the inherent oscillation of the electric field. Throughout the development of optics, every breakthrough regarding the interaction between these three natures has not only deepened our knowledge but also opened a new era of optics, causing the field to flourish. The renowned Malus's law demonstrates that modulating the polarization of a light beam can be turned into amplitude adjustment with the aid of a polarizer, thereby establishing the polarization-to-amplitude link. Academically speaking, the polarization-to-amplitude link has deepened our understanding of the behavior of light beams.

The mutual relationship between amplitude and phase is the cornerstone of scalar optics, and it can be established by wavefront modulation techniques [1-4]. In past centuries, the world has witnessed great successes in scalar optics. The rise of optical research areas, including optical communications [5,6], optical storage [7-9], optical displays [10,11], optical lithography [12], and optical imaging [13-15], have had huge impacts on our lives. However, at our current stage of optical development, the limitations of scalar optics are gradually being realized. For example, the free-space communication capacity based on orbital angular momentum multiplexing can be increased by a factor of two if vector polarized modes are employed [16, 17]. By focusing a cylindrical vector beam, the diffraction limit can be exceeded; however, that is considered to be impossible in the framework of scalar optics [18,19]. Lithography polarization optimization has become a core technique for enhancing the resolution of optical lithography [20,21]. Undoubtedly, the future will forge ahead from scalar optics into an era of vector optics; we are currently only at the outset of this shift. The key to fully entering this era requires a polarized-spatial light modulator (polarized-SLM) that can modify the polarization state of a light beam at a resolution down to the single-pixel level dynamically and in real time.

The polarization-to-phase link demonstrated by the Pancharatnam-Berry (PB) phase in 1987 ushered in the hope of successfully addressing this issue [22]. A polarization convertor based on the PB phase can convert a right-hand circularly polarized beam into a left-hand circularly polarized one along with an additional phase and vice versa. Therefore, a spatially variant linearly polarized beam can be obtained by passing the polarization convertor through a linearly polarized beam. Because of the potential of phase and polarization manipulation, the polarization-to-phase link has received special attention and inspired research on metasurfaces [23-25] and Q-plates [26-28]. However, due to the fixed structures of polarization convertors, dynamic polarization manipulations have thus far been considered to be impossible. Worse, the principle of the PB phase, in essence, hinders the coexistence of circular and linear polarization within a single complete light beam simultaneously, thereby making full polarization modulation impossible. Both of these flaws imply that the polarization-to-phase link cannot meet the requirements for a polarized-SLM.

According to classical optics, of the above three pairs of relationships, vector polarization is linked to scalar phase and amplitude, but not vice versa. However, if the inverse correlation of the polarization-to-phase link could be established, the scalar phase of the light beam could be entirely vectorized into its vector polarization. The adjustment of polarization would no longer be dependent on the structure of the polarized convertor but rely only on the phase of the light beam. In this case, pixelate devices such as phase-only SLM can be utilized to form a polarized-SLM, thus enabling pixel-level polarization modulation of the light beam in a real-time dynamic way. In this context, establishing the phase-to-polarization link is an effective way to tackle both of the major flaws of the polarization-to-phase link; however, due to the entanglement of the inherent polarized modes within one identical light beam, the fundamental phase-to-polarization link is still far beyond our understanding of the vector polarization and scalar phase of the light beam.

Here, we establish the phase-to-polarization link through the principle of phase vectorization. As the inverse process of the PB phase, phase vectorization is achieved by extracting the inherent desired polarized mode from a light beam with different polarization responses using a filter system. Based on phase vectorization, the polarization state of the light beam is determined merely by the scalar phase. Thus, polarized-SLM based on phase vectorization can not only achieve a fully polarized modulation in a real-time dynamic way but also retain complete phase control of the light beam, which is considered to be the key for vector optics.

## Results

### Phase vectorization using a vortex vector beam

Phase vectorization represents the phase-to-polarization link indicated by arrow D in Supplementary Figure 1, and it is capable of vectorizing the scalar phase of the light beam into its vector polarization. As demonstrated in Supplementary Note 1, there are two critical conditions to achieve phase vectorization: one is that the light beam must possess inherently different polarization responses from the left and right circularly polarized modes; the other is that the undesired circularly polarized mode must be eliminated without affecting the desired mode. Normally, both the left and right circularly polarized modes in Supplementary Equations (2) and (3) are entangled during propagation in free space; consequently, extracting only one of the polarized modes from a light beam has always been thought to be a hopeless task.

Although a direct separation of both circularly polarized modes within a light beam cannot be achieved, an indirect solution can be realized by focusing a suitable vector light beam. Specifically, some vector light beams can be divided into left and right circularly polarized modes with different positions in the focal region after focusing by an objective lens. Using a filter system, one of the polarized modes can easily be obtained by eliminating the other one, thereby permitting the realization of phase vectorization. Here, we take the vortex vector beam (VVB) as an example to vectorize the scalar phase into vector polarization, which can be considered as a special case of the generalized phase vectorization depicted in Supplementary Note 1. In this case, $f(\varphi,\theta){=}m\varphi$ in Supplementary Equation (1), and the light beam turns into an $m$-order VVB, which can be written as [16]

$$\mathbf{E}_v = \exp(im\varphi)|\mathbf{R}\rangle + \exp(-im\varphi)|\mathbf{L}\rangle, \tag{1}$$

where $|\mathbf{L}\rangle{=}[1 \quad i]$ and $|\mathbf{R}\rangle{=}[1 \quad -i]$ denote the left and right circularly polarized modes, respectively, and $\varphi$ is the azimuthal angle. When modulating by the phase $\phi{=}\pm(m\varphi-\beta)$, the $m$-order VVB in Equation (1) is transformed into

$$\mathbf{E}_{vl} = \exp[i(2m\varphi-\beta)]|\mathbf{R}\rangle + \exp(-i\beta)|\mathbf{L}\rangle, \tag{2}$$

$$\mathbf{E}_{vr} = \exp[-i(2m\varphi-\beta)]|\mathbf{L}\rangle + \exp(i\beta)|\mathbf{R}\rangle. \tag{3}$$

Here, we call the first and second terms in Equations (2) and (3) undesired and desired polarized modes, respectively. After focusing by an objective lens, the undesired polarized modes $\exp[i(2m\varphi-\beta)]|\mathbf{R}\rangle$ and $\exp[-i(2m\varphi-\beta)]|\mathbf{L}\rangle$ are located at the periphery of the focal plane, while the desired modes $\exp(-i\beta)|\mathbf{L}\rangle$ and $\exp(i\beta)|\mathbf{R}\rangle$ are at the center [see Supplementary Figure 5]. The larger order of $m$ is, the larger the distance between the undesired and desired polarized modes is. After the distance is sufficiently large, the undesired polarized modes can be eliminated using a pinhole to retain only the desired modes. Thus, the polarized modes after the pinhole can be simplified to

$$\mathbf{E}_{pl} = \exp(-i\beta)|\mathbf{L}\rangle, \tag{4}$$

$$\mathbf{E}_{pr} = \exp(i\beta)|\mathbf{R}\rangle. \tag{5}$$

Equations (4) and (5) demonstrate that the polarized modes $\mathbf{E}_{pl}$ and $\mathbf{E}_{pr}$ are determined merely by the phase $\phi = \pm(m\varphi - \beta)$. After reconstruction by an objective lens, both the reconstruction polarizations, $\mathbf{E}_{pl}$ and $\mathbf{E}_{pr}$, link directly with the VVB phase [See Supplementary Note 2: Part 2]. Without loss of generality, one can simply extend this link into a generalized form, namely, phase vectorization [see Supplementary Note 1]. That is, the phase in Equation (6) can be vectorized into the vector polarization in Equation (7):

$$\phi = \text{Phase}[\cos\varphi_0 \exp[i(m\varphi - \beta)] + \sin\varphi_0 \exp[-i(m\varphi - \beta)]]. \tag{6}$$

$$\mathbf{E} = \cos\varphi_0 \exp(-i\beta)|\mathbf{L}\rangle + \sin\varphi_0 \exp(i\beta)|\mathbf{R}\rangle. \tag{7}$$

In Equation (6), the parameter $\varphi_0$ is a weight factor that adjusts the proportion of $|\mathbf{L}\rangle$ and $|\mathbf{R}\rangle$. As shown in Figure 1, the phases $\phi_{1,2} = \pm(m\varphi - \beta)$ can be vectorized into left and right circular polarizations, respectively. For linear polarization, there is a one-to-one correspondence between the parameter $\beta$ of $\phi_3 = \text{Phase}[\cos(m\varphi - \beta)]$ and the polarization direction. Thus, a radially polarized beam is easily achieved by the phase $\phi_3$ with $\beta = \varphi$. Note that the phases $\phi_{1,2} = \pm(m\varphi - \beta)$ are obtained by $\varphi_0 = 0$ and $0.5\pi$ in Equation (6), while the $\phi_3 = \text{Phase}[\cos(m\varphi - \beta)]$ is the result of Equation (6), where $\varphi_0 = 0.25\pi$. For other $\varphi_0$ values, the phase in Equation (6) can be vectorized into an elliptical polarization accordingly.

**Polarized-SLM based on phase vectorization**

In the following experiment, we establish a polarized-SLM based on phase vectorization using $m = 30$-order VVB. Figure 2 presents a schematic of polarized-SLM (a simplified version is shown in Supplementary Figure 4), where the polarized-SLM is composed of a phase control system (the green dotted box) and a filter system (the blue dotted box). In the phase control system, a collimated incident x linearly polarized beam with a wavelength of 633 nm propagating along the optical axis passes through a phase-only SLM and two lenses (L₃, L₄) before being converted into a $m$=30-order VVB by a vortex polarizer (VP). The VP can be easily manufactured using the Q-plate technique [27, 28]. Figure 2(a and b) shows the light intensities of $m$=30-order VVB without and with a polarizer, as indicated by the purple arrow. L₃ ($f_3$=150 mm) and L₄ ($f_4$=150 mm) compose a 4$f$-system that causes the phase coded in the phase-only SLM and VP to conjugate. Thus, the modulated VVB in Equations (2) and (3) can easily be obtained by coding the phase $\phi = \pm(m\varphi - \beta)$ in the phase-only SLM, where $m$=30.

In the latter system, the modulated VVB in Equations (2) and (3) is divided into left and right circularly polarized modes with different positions in the focal region after focusing by the objective lens, OL₁. Specifically, the undesired polarized modes $\exp\left[i(2m\varphi - \beta)\right]|\mathbf{R}\rangle$ and $\exp\left[-i(2m\varphi - \beta)\right]|\mathbf{L}\rangle$ are located at the outer ring and have a topologic charge of $\pm 60$, while the desired polarized modes $\exp(-i\beta)|\mathbf{L}\rangle$ and $\exp(i\beta)|\mathbf{R}\rangle$ are in the geometric focal position of OL₁ [see Supplementary Note 2]. The distance between both pairs of polarized modes is 671 μm. After passing through a pinhole with a radius of 400 μm, the undesired polarized modes are eliminated, and the desired modes can be further recovered by a subsequent objective lens, OL₂. Because the desired polarized modes in the focal region of OL₁ depend only on the VVB phase, the polarization from OL₂ also possesses a direct link with the VVB phase. That is, the vector polarization can be manipulated entirely at will by merely changing the VVB phase. Here, the numerical apertures (NA) of both OL₁ and OL₂ are 0.01, and the red arrow represents the propagation direction of the light beam.

Figure 3 presents the experimental results of the phase vectorization shown in Figure 1 (the theoretical results are shown in Supplementary Figure 6). The phases $\phi = \pm(m\varphi - \beta)$ with $m$=30 and $\beta = 0$ in Figure 3(a and e) are vectorized into left and right circular polarizations, respectively, which can be distinguished from each other using a quarter-wave plate (indicated by the blue arrow) and a polarizer (indicated by the purple arrows), as shown in Figure 3(b, c, d, f, g, and h). In terms of linear polarization, a light beam with an on-demand polarization direction can be achieved by altering the phase $\phi = \text{Phase}[\cos(m\varphi - \beta)]$. After passing through a horizontal polarizer (purple arrow), different polarization directions lead to different light intensities, which are determined by the parameter $\beta$. Here, $\beta$=0, 0.25$\pi$, 0.5$\pi$, and 0.75$\pi$ in Figure 3(m, n, o, and p), respectively. The varieties of light intensities in Figure 3(i, j, k, l) imply that the polarization direction has a one-to-one correspondence with the parameter $\beta$, as shown in Figure 1. Thus, radial polarization is realized with $\beta=\varphi$, see Supplementary Figure 10 (a–c). Although it is beyond imagination, the inverse process of the PB phase, namely, phase vectorization, can be realized using high-order VVB. Note that the hollow shape of the vector beam from OL$_2$ is attributable to the slight deviation between the conjugated planes of the phase-only SLM and VP; however, that does not affect the validity of the polarized-SLM.

Figure 3 not only verifies the full polarization modulation of polarized-SLM but also implies the essence of phase vectorization. According to Equations (6) and (7), linear polarization can be obtained by the phase $\phi = \mathrm{Phase}[\cos(m\varphi - \beta)]$, where the parameter $\beta$ controls the polarization direction. As shown in Figure 3(m, n, o, and p), all the phases for the creation of linear polarization have a single identical binary phase structure determined by the $m$-order of the VVB. However, different $\beta$ values induce a relative phase displacement in Figure 3(m, n, o, and p), which causes a further polarization change in Figure 3(i, j, k, l). Similarly, left and right circular polarizations correspond to the vortex phases $\phi = \pm(m\varphi - \beta)$, respectively. Both vortex phases have the same phase structure but an inverse topological charge of $\pm m$, see in Figure 3(a and e). In this case, the modification of left and right circular polarization is no longer dependent on the parameter $\beta$ but on the sign of the topological charge $\pm m$. The relative displacement of the phase structure caused by $\beta$ only imposes an additional phase on circular polarization. Thus, the essence of phase vectorization is to transfer the polarization change into three specific phases: binary phase $\phi = \mathrm{Phase}[\cos(m\varphi - \beta)]$ for linear polarization, vortex phase $\phi = \pm(m\varphi - \beta)$ for circular polarization, and a combination of both for elliptical polarization; thus, elliptical polarization is a combination of circular and linear polarization.

**Pixel-level modulation using polarized-SLM**

Phase vectorization establishes the phase-to-polarization link that enables polarization to be manipulated directly through the VVB phase. Generally, the above three particular phases of VVB can be pixelized by the phase-only SLM in Figure 2. The pixelate phase enables pixelate polarization, thereby permitting individual adjustment of the polarization in each pixel, as shown in Figure 4. Here, one example of Figure 4 is presented in Supplementary Figure 7.

Make no mistake, the pixel size of output vector beam in Figure 4 is not equal to that of phase-SLM in Figure 2. Because the desired polarized mode is passing through the filter system in Figure 2, the resolution of polarized-SLM is lower than that of phase-SLM. Generally, the smaller the pixel of phase-SLM, the higher the resolution of polarized-SLM in Figure 2. As the development of phase-SLM, the size of pixel will become sufficiently small in future, thereby enhancing the resolution of polarized-SLM accordingly. However, this hardware problem does not affect the validity of the pixelate polarization of polarized-SLM in Figure 4.

To verify the pixel-level polarization modulation of polarized-SLM, Figure 5 presents the experimental result of performing pixel-level polarization modulation using polarized-SLM. As shown in Figure 5(a, d, and g), the VVB phases possess four pixel-like zones (A, B, C, and D). Their corresponding polarization states are shown in Figure 5(j, k, l, and m), respectively, and can be expressed as follows:

$$P_n = \begin{bmatrix} \cos\left[n\varphi + (n-1)\pi/4\right] \\ \sin\left[n\varphi + (n-1)\pi/4\right] \end{bmatrix}. \tag{8}$$

where $n$=1, 2, 3, and 4 represent zones A, B, C, and D, respectively. The backgrounds of the light beams in Figure 5(b, e, and h) are linear polarizations adjusted to be orthogonal to the polarizer by the parameter $\beta$ of the phase $\phi = \mathrm{Phase}[\cos(m\varphi - \beta)]$. Here, the polarization directions of the background light beams (yellow arrows) in Figure 5(c, f, and i) are $0.5\pi$, $0.75\pi$, and $0$, respectively, and their corresponding polarizers (purple arrows) are $0$, $0.25\pi$, and $0.5\pi$, respectively. Owing to the orthogonality between the background light beams and the polarizers, the background light beams are always eliminated by the polarizers, thereby providing a clear display of the petal-like patterns in zones A, B, C, and D. As shown in Figure 5(c, f, i), the petals with $2n$ numbers are rotated along with the polarizer in zones A, B, C, and D, which is coincident with the theoretical results in zones A, B, C, and D in Supplementary Figure 9. In addition, we generate other vector beams in Supplementary Note 3 by patterning the phase in polarized-SLM.

## Discussion

### Judgment of the phase-to-polarization link

The key of this work lays on whether the fundamental phase-to-polarization is established or not. As shown in Supplementary Figure 1, the phase and polarization both come from one identical vector beam. Thus, there are two criterions that can judge whether this fundamental link is established or not.

(1) Whether the light beam is satisfied Helmholtz equation. Light beam satisfied Helmholtz equation can propagate in free space, which is normally considered as a light beam in optics.

(2) Whether the phase in Equation (6) and the polarization in Equation (7) are the phase and polarization of light beam, respectively.

Mathematically, this fundamental link can finally be derived into Equation (6) and Equation (7). That is, the phase in Equation (6) can link with the polarization in Equation (7). For the first criterion: $m$-order VVB is a special solution of Helmholtz equation, which is widely used in optics. Thus, $m$-order VVB is a light beam that can propagate in free space. For the second criterion: phase vectorization is a process of inherent polarized mode extraction for light beams. That is, the desired and undesired polarized modes come from the $m$=30-order VVB, which further demonstrates that the polarization in Equation (7) output from the polarized-SLM is the inherent polarized mode of $m$-order VVB; because the phase of VVB can be adjusted at will by the phase-only SLM in Figure 2, the phase coded in the phase-only SLM is the phase of $m$-order VVB. Based on both above criterions, the phase-to-polarization link is established in classical optics. It should be emphasized that this fundamental link is generalized. As shown in Figure 6, we only present the first example of the phase-to-polarization link using VVB. If only light beam satisfies the two conditions in Supplementary note 1, one can always establish this fundamental link.

**New principle of polarization modulation**

According to the above judgment criterions, the phase-to-polarization link is totally different from the conventional principle of polarization modulation. Generally, polarization manipulation always requires two orthogonally polarized beams with controllable phases. For example, one can achieve full polarization adjustment by using a metasurface with effective birefringence, where the phases of two electric components $\mathbf{E}_x$ and $\mathbf{E}_y$ can be adjusted at will [23]. Using an interferometric configuration [29, 30], a light beam with arbitrary polarization can be achieved by superposing left and right circular polarized beams with different phases. Unlike this conventional principle, phase vectorization is not the result of superposing two orthogonally polarized light beams but a direct control on the electric field of one identical light beam by vectorizing the scalar phase into its vector polarization [See Supplementary Note 4]. Thus, polarized-SLM not only does not require a complicated algorithm, high interferometrically precise alignment, or an expensive and difficult fabrication process, and it can achieve the dynamic and real-time adjustment of the polarization of a light beam. From this perspective, the principle of phase vectorization provides entirely new knowledge regarding polarization manipulation.

**Theoretical impact of phase vectorization**

In this section, we discuss the theoretical impact of phase vectorization in classical optics. Heretofore, there have been three relationships between amplitude, phase and polarization in classical optics, namely, the mutual link between phase and amplitude, the polarization-to-amplitude link and the polarization-to-phase link. These three links imply that vector polarization can link to the scalar phase and amplitude, but not the opposite. That is, these three properties of the light beam are relatively independent. Phase vectorization establishes a phase-to-polarization link by transforming all the polarization states into three specific phases. By combining the above three relationships with the principle of phase vectorization, one can predict that the phase, amplitude and polarization of the light beam can be adjusted simultaneously merely by the phase. Thus, the phase-to-polarization link can be considered as the last step in unifying amplitude, phase and polarization. For more additional discussions, please refer to Supplementary Note 4.

In conclusion, we have theoretically and experimentally demonstrated the phase-to-polarization link by the principle of phase vectorization. Phase vectorization realized by the phase modulation of $m$=30-order VVB in a filter system can vectorize three specific phases—the binary phase, vortex phases, and the combination of both—into linear, circular and elliptical polarizations, respectively. Using a phase-only SLM, polarized-SLM is established based on phase vectorization, enabling both pixel-level polarization and phase manipulation of the light beam dynamically and in real time. In theory, phase vectorization not only extends our understanding of the relationship between the scalar phase and vector polarization but also makes it possible to manipulate amplitude, phase and polarization merely by the phase, opening a new avenue for the development of full-SLM. In addition, and as key aspect for entering the era of vector optics, real-time pixel-level polarization modulation using polarized-SLM offers promising applications in scientific fields such as optical communication [17] and optical lithography [20, 21].

## Data Availability

All data supporting the findings of this study are available from the corresponding author on request.

**Acknowledgments**

Parts of this work were supported by the National Natural Science Foundation of China (61525503/61620106016/11804232); the National Key Research and Development Program of China (2018YFC1313803).

**Author contributions**

X. Weng conceived of the research and designed the experiments. G. Sui, X. Dong and X. Weng performed the experiments. Q. Song and X. Weng analyzed all of the data. X. Weng and X. Gao co-wrote the paper, and J. Qu offered advice regarding its development. X. Gao, J. Qu and S. Zhuang directed the entire project. All authors discussed the results and contributed to the manuscript.

**Competing interests statement**

A provisional patent application (PCT/CN2020/082764) has been filed on the subject of this work.

# Figures

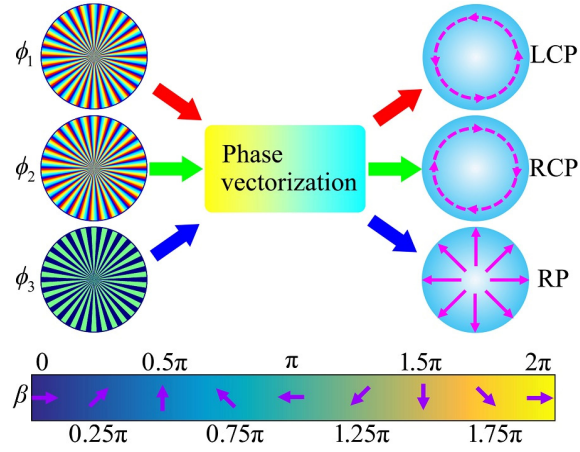

**Figure 1.** Schematic of phase vectorization. The phase $\phi_{1,2} = \pm(m\varphi - \beta)$ can be vectorized into left and right circular polarizations, respectively, while linear polarization can be obtained by the phase $\phi_3 = \text{Phase}[\cos(m\varphi - \beta)]$, where $\beta$ indicates the polarization direction. For example, a radially polarized beam can be obtained by the phase $\phi_3$ with $\beta = \varphi$.

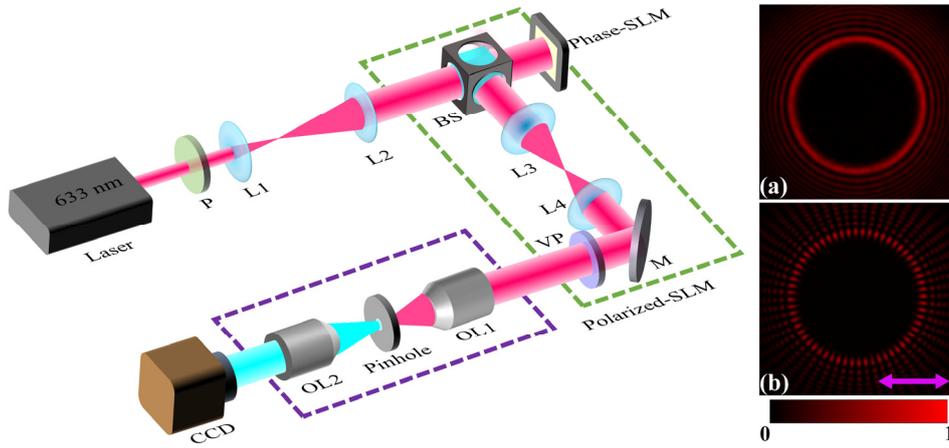

**Figure 2.** Polarized-SLM based on phase vectorization. The phase control system (green dotted box) and the filter system (violet dotted box) compose an entire polarized-SLM. $L_3$ ($f_3$=150 mm) and $L_4$ ($f_4$=150 mm) compose a 4f-system that causes the phase coded in the phase-only SLM and VP to conjugate. The light beam modulated by the phase-only SLM is converted into an $m$=30-order VVB by a vortex polarizer (VP). Subfigures (a) and (b) show the light intensities of $m$=30-order VVB without and with a polarizer (purple arrow), respectively. The modulated VVB output from the first system is divided into a desired and an undesired polarized mode. After passing through the filter system, the undesired mode is eliminated by the pinhole (PH) in the focal region of the objective lens $OL_1$, and only the desired mode is retained. After being reconstructed by the objective lens, $OL_2$, the desired polarized mode is transformed into the desired polarization, which is recorded using a CCD.

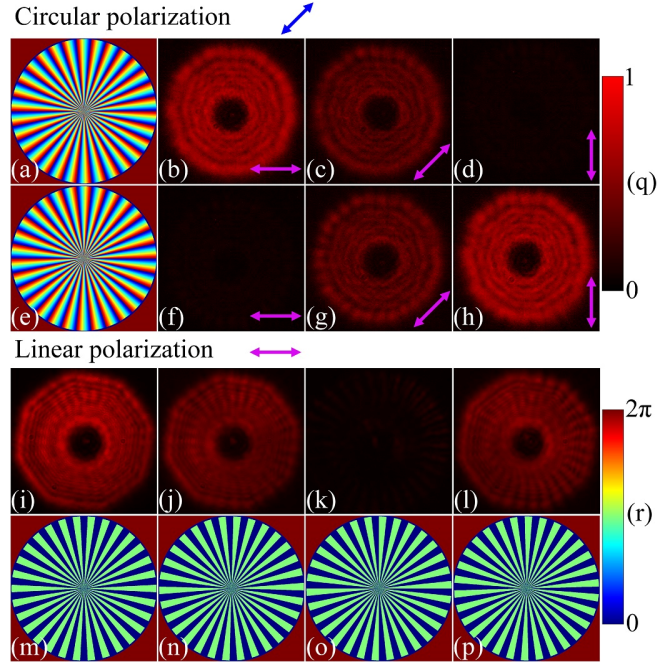

**Figure 3.** Experimental results of the phase vectorization in Figure 1. The vortex phases $\phi = \pm(m\varphi - \beta)$ with $m$=30 and $\beta$=0 (a, e) are vectorized into left and right circular polarizations, respectively. Both circular polarizations are analyzed by observing the light intensities (b, c, d, f, g, and h) passing through a quarter waveplate. The fast axis is indicated by the blue arrow and the polarizers are indicated by the purple arrows. Linear polarization is achieved by the phase $\phi = \mathrm{Phase}[\cos(m\varphi - \beta)]$, where $\beta$=0, $0.25\pi$, $0.5\pi$, and $0.75\pi$ in (m, n, o, p), respectively. After passing through a polarizer (purple arrow), linear polarizations in different directions exhibit different amplitudes of light intensity (i, j, k, l). The intensities of all the light beams are normalized to a unit value, as indicated by color bar (q). Color bar (r) shows the phase scales of (a, e, and m–p).

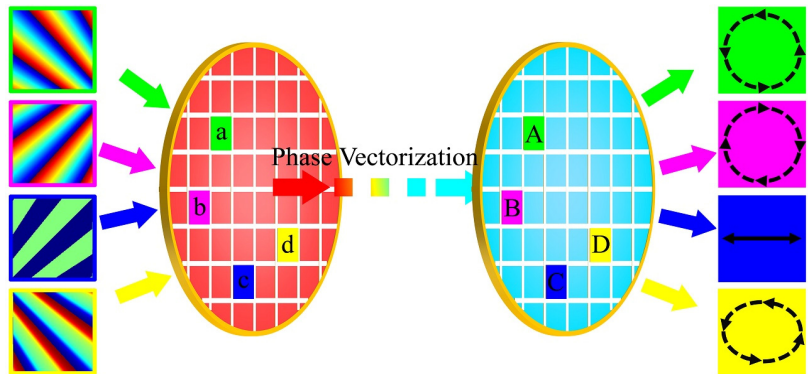

**Figure 4.** Schematic of pixelate polarization modulation achieved by the pixelate phase. The letters a, b, c, and d denote the four phases of the light beam. Their corresponding polarizations of A, B, C, and D, are indicated by the black arrows.

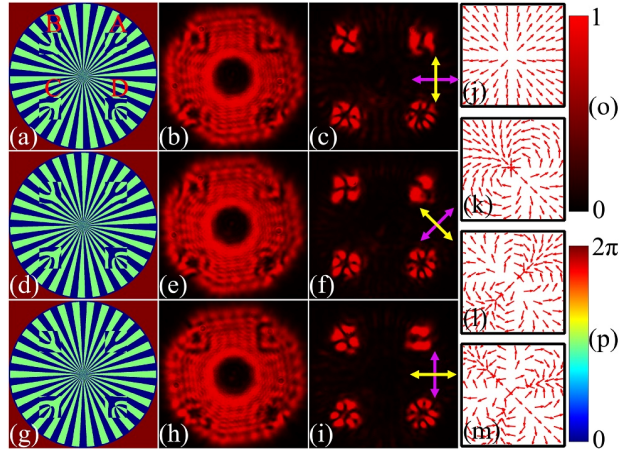

**Figure 5.** Pixel-level polarization modulation using polarized-SLM. The light beams (b, e, and h) created by phases (a, d, and g), respectively, are composed of background and four pixel-like zones, namely, zones A, B, C, and D. The polarizations of zones A, B, C, and D are shown in (j, k, l, and m), respectively and can be expressed as in Equation (8). The background polarizations (yellow arrows) are adjusted to be orthogonal to the polarizers (purple arrows). When passing through the polarizers, the background light beams are always eliminated, thereby providing the clear petal-like pattern displays in zones A, B, C, and D (c, f, and i). The intensities of all the light beams are normalized to a unit value as indicated by color bar (o). Color bar (p) shows the phase scales of (a, d, and g).

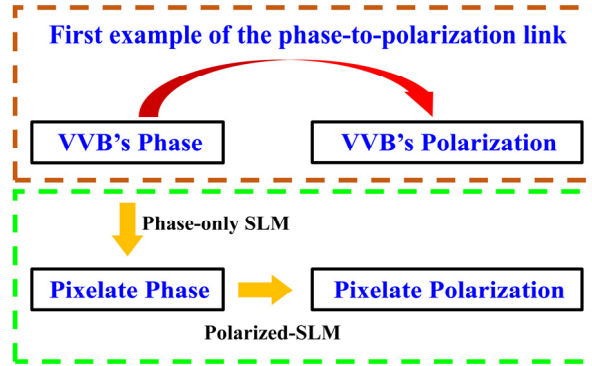

**Figure 6.** First example of the phase-to-polarization link using VVB. Because the VVB's phase can be pixelated by phase-only SLM, pixelate polarization can therefore be achieved by the polarized-SLM in Figure 2.

# Supplementary Information

## *Establishment of the fundamental phase-to-polarization link in classical optics*

Weng et al.

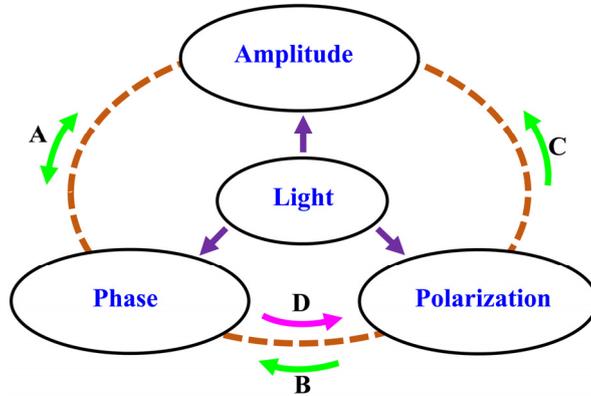

**Supplementary Figure 1.** The relationship between amplitude, phase and polarization of light: arrow A denotes the mutual relationship between amplitude and phase; arrow B indicates the polarization-to-phase link; and arrow C denotes the polarization-to-amplitude link. Arrow D represents the phase-to-polarization link.

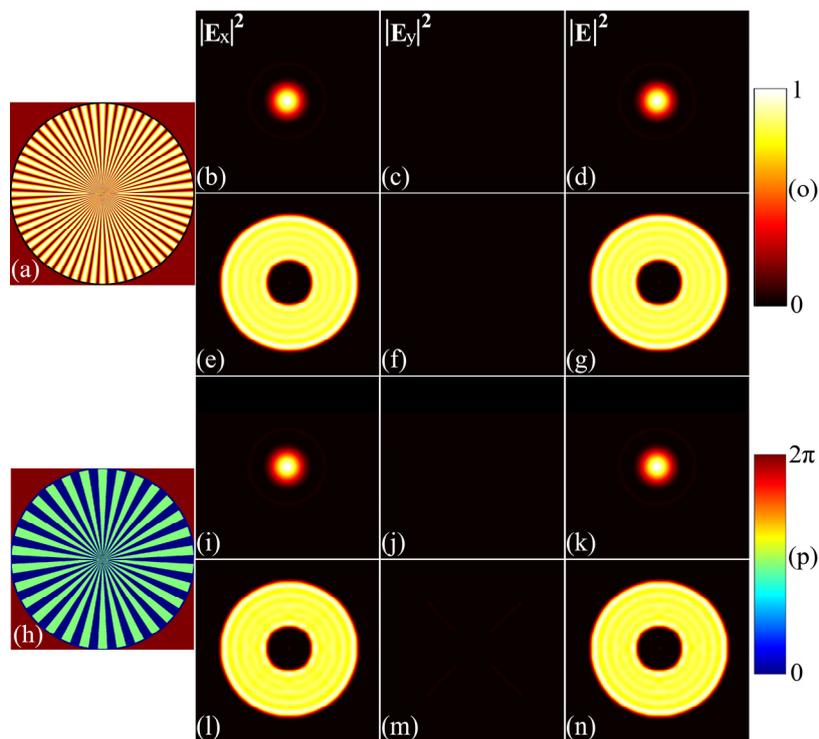

**Supplementary Figure 2.** Linearly polarized beams generated by accurate and approximate modulation. The light beams (g and n) are created by vectorizing the accurate and approximate wavefronts in (a and h), respectively. Subfigures (e, f, l, and m) show the light intensities passing through a polarizer with horizontal polarized direction. Subfigures (b–d and i–k) show the x component, y component and total light intensities of the corresponding desired polarized modes in the focal region of $OL_1$. The intensities of all the light beams are normalized to a unit value, as indicated by the color bar (o). The color bar (p) shows the phase scales of (a, d, and g).

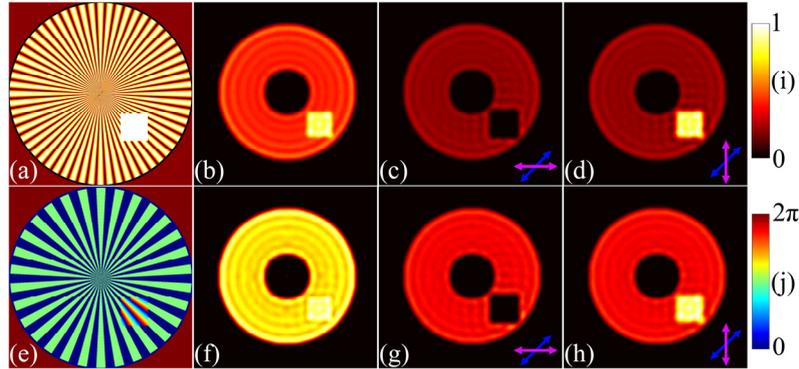

**Supplementary Figure 3.** Special vector beams generated by accurate and approximate modulation. The light beams (b and f) are composed of x-linear polarizations in the pixel-like zone and right circular polarizations in the background created by vectorizing the accurate and approximate wavefronts in (a and e), respectively. Subfigures (c, d, g, and h) show the light intensities passing through the polarizer (purple arrow) and quarter waveplate (blue arrow); these intensities are normalized to a unit value, as indicated by the color bar (i). Color bar (j) shows the phase scales of (a and e).

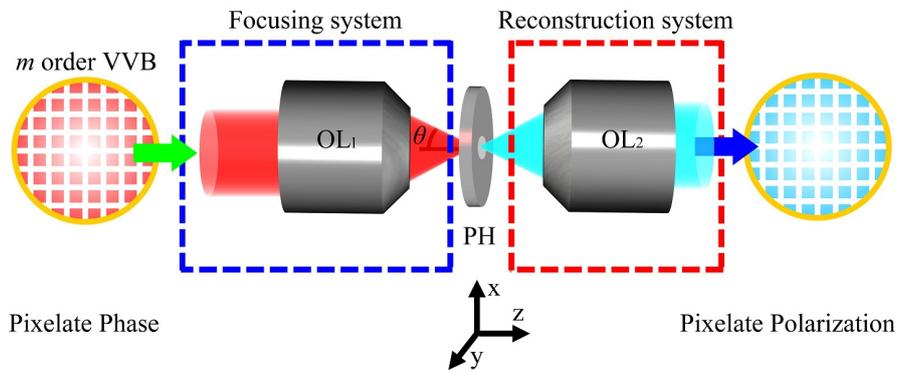

**Supplementary Figure 4.** Simplification of polarized-SLM based on phase vectorization. Polarized-SLM can be simplified into a filter system composed of a focusing system and a reconstruction system. After passing through polarized-SLM, the pixelate phase of $m$-order VVB can be vectorized into pixelate polarization. PH denotes a pinhole, and $OL_1$ and $OL_2$ are two objective lenses, both with a numerical aperture of 0.01. $\theta$ is the convergent angle of $OL_1$.

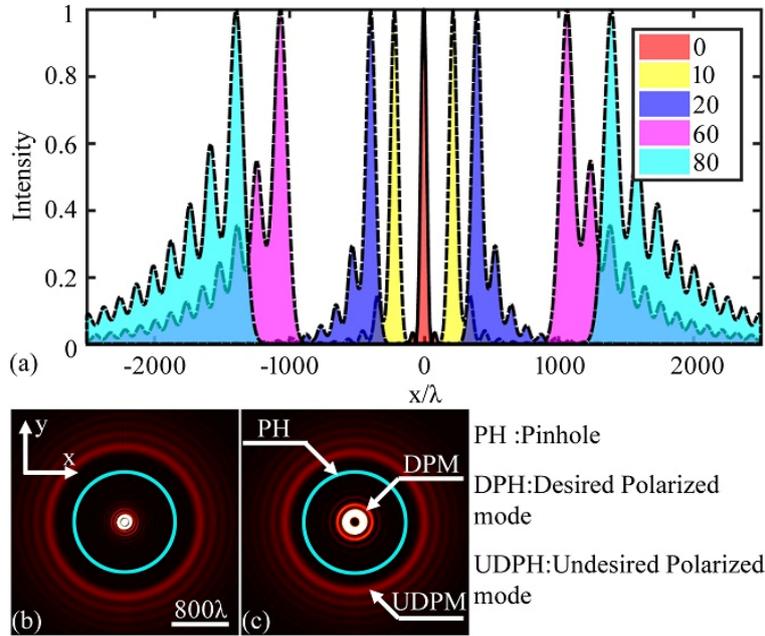

**Supplementary Figure 5.** Spatial separation of desired and undesired polarized modes in the focal region of OL₁. (a) The light intensities of the polarized modes with different topological charges along the x axis; (b, c) The focal light intensity of the VVB obtained by the modulation of the phase $\phi = m\varphi - \beta$ with $m$=30; (b) $\beta = 0$; (c) $\beta = 5\varphi$. DPM: Desired polarized mode located at the center; UDPM: Undesired polarized mode located at the outer ring; PH: pinhole placed between DPM and UDPM.

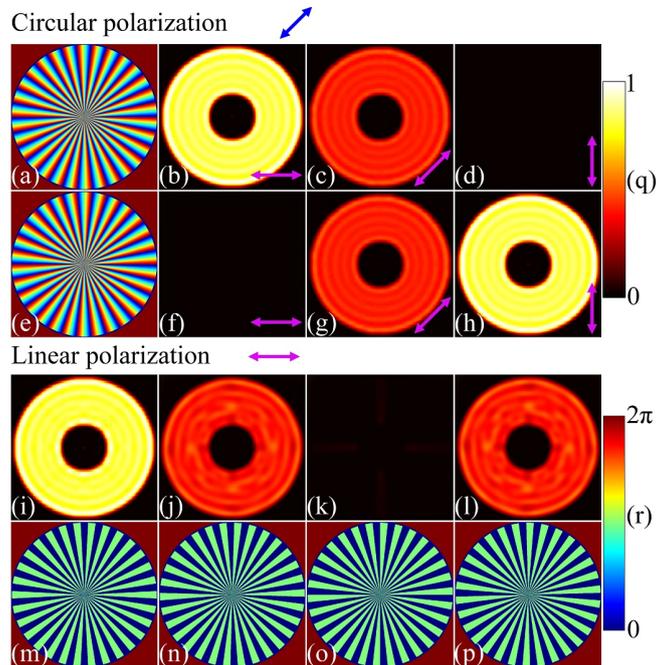

**Supplementary Figure 6.** Theoretical results of Figure 3. The vortex phases $\phi = \pm(m\varphi - \beta)$ with $m$=30 and $\beta$=0 (a and e) are vectorized into left and right circular polarizations, respectively, while the linear polarizations are realized by the phases $\phi = \text{Phase}[\cos(m\varphi - \beta)]$ with $\beta$=0, 0.25$\pi$, 0.5$\pi$, 0.75$\pi$ in (m, n, o, and p), respectively. Subfigures (b–d, f–h, and i–l) show the light intensities passing through the polarizer (purple arrow) and quarter waveplate (blue arrow) and are normalized to a unit value as indicated by color bar (q). Color bar (r) shows the phase scales of (a, e, and m–p).

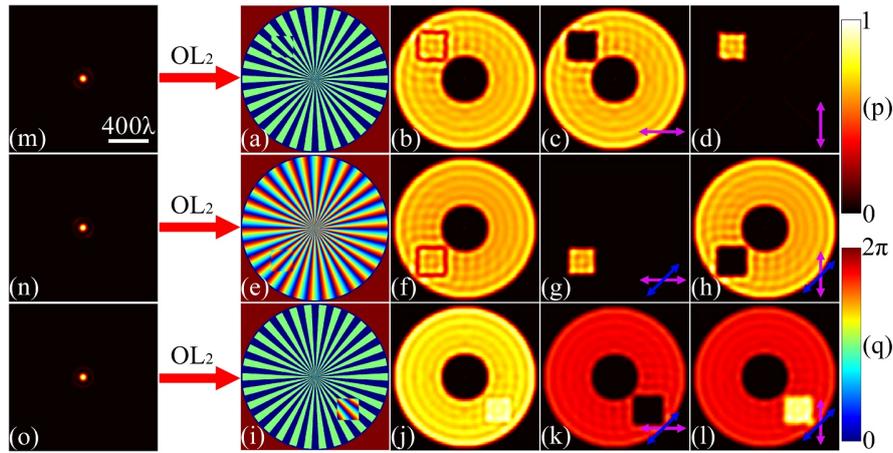

**Supplementary Figure 7.** Polarization modulation in an arbitrary pixel-like zone of a light beam. The light beams in (b, f, and j) are created by the phases in (a, e, and i), respectively. Subfigures (m, n, and o) show the desired polarized modes in the focal region of $OL_1$ that are further reconstructed by $OL_2$ to form the desired polarizations in (b, f, and j), respectively. Subfigures (c, d, g, h, k, and l) show the light intensities of the light beams in (b, f, and j) passing through the polarizer (indicated by the purple arrow) and the quarter waveplate (indicated by the blue arrow). The intensities of all the light beams are normalized to a unit value as indicated by color bar (p). Color bar (q) shows the phase scales of (a, e, and i).

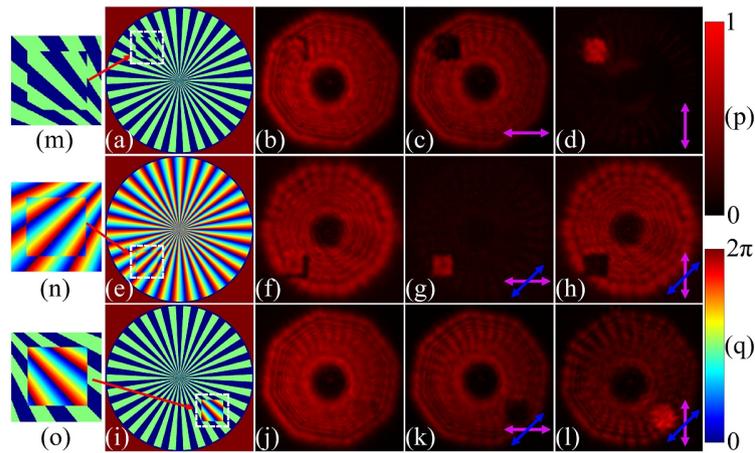

**Supplementary Figure 8.** Theoretical results of Supplementary Figure 7. The light beams (b, f, and j) are composed of a pixel-like zone and background created by vectorizing the phase in (a, e, and i), respectively. By observing the light intensities (c, d, g, h, k, l) passing through the polarizer (purple arrow) and the quarter waveplate (blue arrow), the polarization states of the background and pixel-like zones are (b) x and y linear polarization; (f) right and left circular polarization; (j) x linear and right circular polarization, respectively. The intensities of all the light beams are normalized to a unit value as indicated by color bar (p). Color bar (q) shows the phase scales of (a, e, and i).

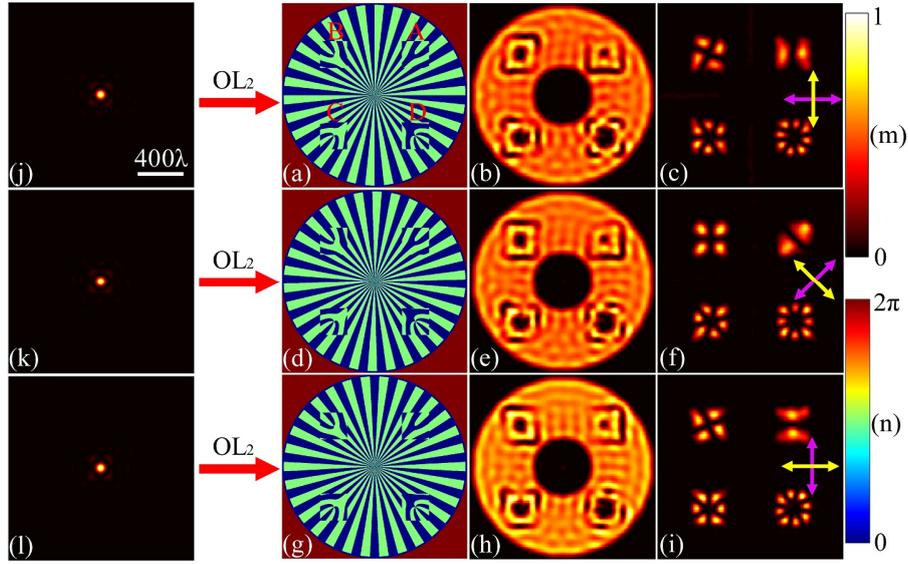

**Supplementary Figure 9.** Theoretical results of Figure 5. The light beams in (b, e, and h) are created by the phases in (a, d, and g), respectively. Subfigures (j, k, and l) show the desired polarized modes in the focal region of $OL_1$ that are further reconstructed by $OL_2$ to form the desired polarizations in (b, e, and h), respectively. Subfigures (c, f, and i) show the light intensities of the light beams in (b, e, and h) passing through the polarizer (purple arrow). Because the background polarizations (indicated by the yellow arrow) are adjusted to be orthogonal with the polarizers (purple arrow), the light intensities of the background are always eliminated, and the polarizations of the A, B, C, and D zones are demonstrated by the petal-like patterns (see (c, f, and i)). The intensities of all the light beams are normalized to a unit value as indicated by color bar (m). Color bar (n) shows the phase scales of (a, d, and g).

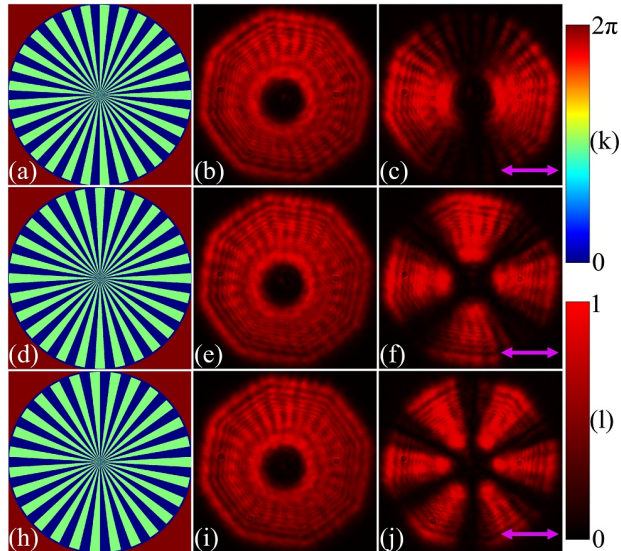

**Supplementary Figure 10.** Arbitrary order VVB. (b) 1st-order VVB; (e) 2nd-order VVB; (i) 3rd-order VVB; (c, f, j) are the corresponding light intensities passing through the polarizer (indicated by the purple arrow). Subfigures (a, d, and h) show the needed phases $\phi = \text{Phase}\{\cos[(m\varphi - \beta)]\}$ with $m=30$; (a) $\beta = \varphi$; (d) $\beta = 2\varphi$; (h) $\beta = 3\varphi$. The intensities of all the light beams are normalized to a unit value as indicated by color bar (l). Color bar (k) shows the phase scales of (a, d, and h).

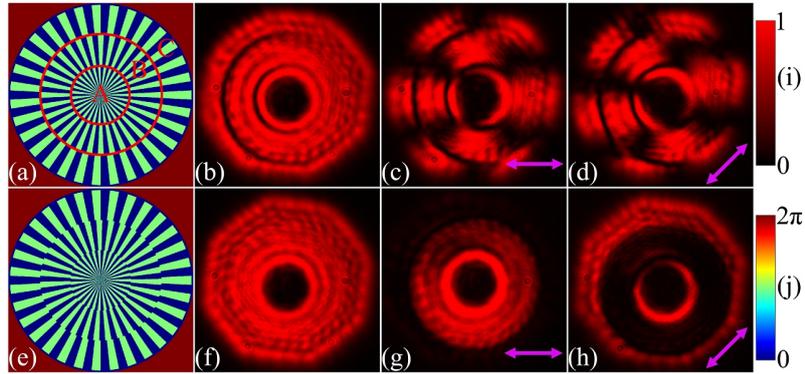

**Supplementary Figure 11.** Vector beam with multiple zones. Subfigures (b, f) show the light beams with three polarized zones, namely, A, B, and C, which are realized by the phases in (a) and (e), respectively. After passing through the polarizers (indicated by the purple arrows), the light intensities in (c, d, g, and h) demonstrate that the polarizations of zone A, B, and C are consistent with those of Supplementary Equations (28) and (29), respectively. The intensities of all the light beam are normalized to a unit value as indicated by color bar (i). Color bar (j) shows the phase scales of (a) and (e).

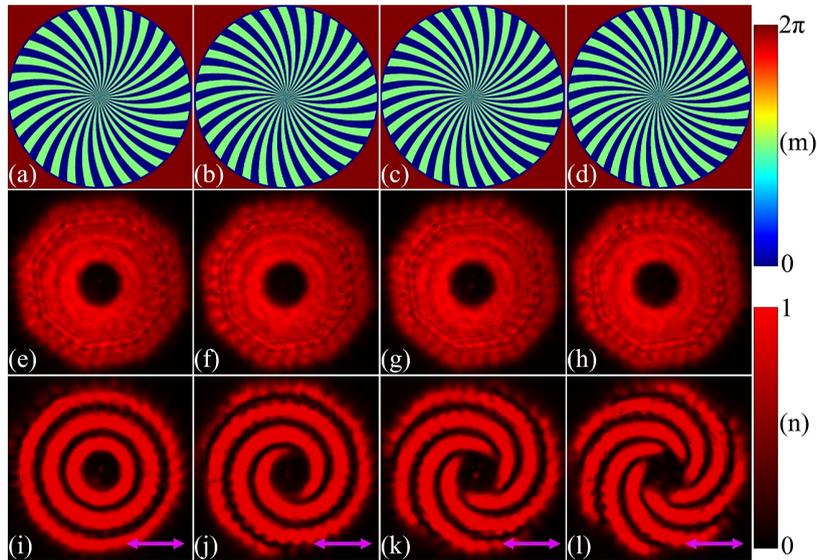

**Supplementary Figure 12.** Special vector beam. The phases (a, b, c, and d) are vectorized into special vector beams (e, f, g, and h), respectively, and (i, j, k, and l) are the corresponding light intensities passing through the polarizers (indicated by the purple arrows). The intensities of all the light beams are normalized to a unit value, as indicated by color bar (n). Color bar (m) shows the phase scales of (a–d).

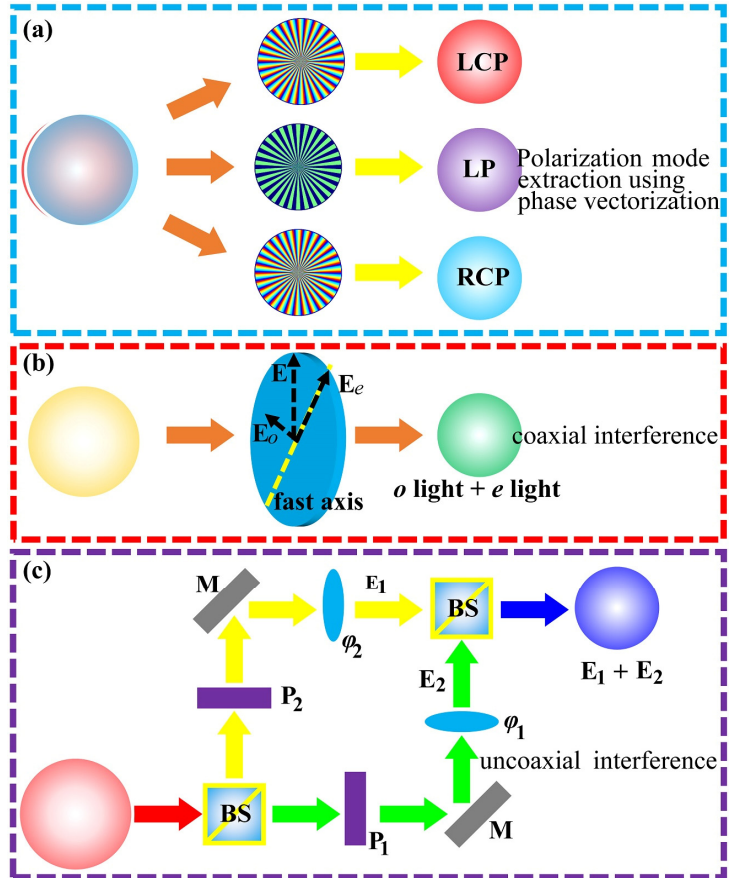

**Supplementary Figure 13.** Different polarization modulation schemes. Subfigures (a, b, and c) are schematics of phase vectorization, the waveplate and the interferometric optical system, respectively. In subfigure (a), LCP, RCP and LP indicate left, right circular, and linear polarization, respectively. In subfigure (b), $\mathbf{E} = \mathbf{E}_o + \mathbf{E}_e$, where $\mathbf{E}_o$ and $\mathbf{E}_e$ are the electric fields of $o$ light and $e$ light, respectively. In subfigure (c), BS: beam splitter; M: mirror; $P_1$, $P_2$: polarizer that converts light beams into two orthogonally polarized beams, namely, $\mathbf{E}_1$ and $\mathbf{E}_2$; and $\varphi_1$, $\varphi_2$: the phases of $\mathbf{E}_1$ and $\mathbf{E}_2$, respectively.

# Supplementary Note 1: The principle of phase vectorization

In classical optics, there are three pairs of relationships between the amplitude, phase and polarization of light beams. As shown in Supplementary Figure 1, arrow A indicates the mutual relationship between amplitude and phase, which are linked with each other through wavefront modulation techniques [1-4]. Malus's law establishes the polarization-to-amplitude link with the aid of a polarizer (denoted by arrow C). The polarization-to-phase link (indicated by arrow B) was demonstrated by the Pancharatnam–Berry (PB) phase in 1987 [5]. From these three links, one can link vector polarization to the scalar phase and to amplitude; however, the reverse relationships do not apply. If the inverse correlation of the PB phase could be established, the scalar phase of the light beam could entirely be vectorized into vector polarization. Here, we call this inverse process of arrow B "phase vectorization," namely, the phase-to-polarization link indicated by arrow D.

## Part 1: Generalized theory

To establish the phase-to-polarization link, the light beam must possess different polarization responses. Such a light beam has a generalized form; its electric field can be expressed as follows:

$$\mathbf{E}_v = \exp\left[if(\varphi,\theta)\right]|\mathbf{R}\rangle + \exp\left[-if(\varphi,\theta)\right]|\mathbf{L}\rangle, \tag{1}$$

where $|\mathbf{L}\rangle = \begin{bmatrix} 1 & i \end{bmatrix}$ and $|\mathbf{R}\rangle = \begin{bmatrix} 1 & -i \end{bmatrix}$ denote the left and right circularly polarized modes, respectively. $f(\varphi,\theta)$ represents a generalized phase function, and $\theta$ and $\varphi$ are the convergence and azimuthal angle of the focusing system in Supplementary Figure 4, respectively.

When modulating by the phase $\phi = \pm\left[f(\theta,\varphi) - \beta\right]$, the light beam in Supplementary Equation (1) is transformed into

$$\mathbf{E}_{vl} = \exp\left[i\left(2f(\theta,\varphi) - \beta\right)\right]|\mathbf{R}\rangle + \exp\left(-i\beta\right)|\mathbf{L}\rangle, \tag{2}$$

$$\mathbf{E}_{vr} = \exp\left[-i\left(2f(\theta,\varphi) - \beta\right)\right]|\mathbf{L}\rangle + \exp\left(i\beta\right)|\mathbf{R}\rangle. \tag{3}$$

Supplementary Equations (2) and (3) imply that the phase change results in different polarization responses of $\exp\left(-i\beta\right)|\mathbf{L}\rangle$ and $\exp\left(i\beta\right)|\mathbf{R}\rangle$ except in the undesired polarized modes $\exp\left[i\left(2f(\theta,\varphi) - \beta\right)\right]|\mathbf{R}\rangle$ and $\exp\left[-i\left(2f(\theta,\varphi) - \beta\right)\right]|\mathbf{L}\rangle$. Supposing that both undesired polarized modes can be eliminated, the light beams in Supplementary Equations (2) and (3) can be simplified to

$$\mathbf{E}_l = \exp\left(-i\beta\right)|\mathbf{L}\rangle, \tag{4}$$

$$\mathbf{E}_r = \exp\left(i\beta\right)\left|\mathbf{R}\right\rangle. \tag{5}$$

According to Supplementary Equations (4) and (5), one can easily obtain the phase-to-polarization link; then, the scalar phase in Supplementary Equation (6) can be vectorized into the vector polarization in Supplementary Equation (7). The reason why we only retain the phase in Supplementary Equation (6) is presented in Part 2.

$$\phi = \text{Phase}[\cos\varphi_0 \exp[i(f(\theta,\varphi)-\beta)] + \sin\varphi_0 \exp[-i(f(\theta,\varphi)-\beta)], \tag{6}$$

$$\mathbf{E} = \cos\varphi_0 \exp\left(-i\beta\right)\left|\mathbf{L}\right\rangle + \sin\varphi_0 \exp\left(i\beta\right)\left|\mathbf{R}\right\rangle. \tag{7}$$

In Supplementary Equation (6), the parameter $\varphi_0$ is a weight factor that adjusts the proportion of $\left|\mathbf{L}\right\rangle$ and $\left|\mathbf{R}\right\rangle$. For $\varphi_0 = 0, 0.5\pi$, $\left|\mathbf{L}\right\rangle$ and $\left|\mathbf{R}\right\rangle$ can be obtained, respectively. For $\varphi_0 = 0.25\pi$, the phase in Supplementary Equation (6) can be simplified as $\phi = \text{Phase}[\cos(f(\theta,\varphi)-\beta)]$, thereby creating a linearly polarized beam with the polarization direction $\beta$, which can be expressed as follows

$$\mathbf{E}_h = \cos\beta\left|\mathbf{x}\right\rangle + \sin\beta\left|\mathbf{y}\right\rangle, \tag{8}$$

where $\left|\mathbf{x}\right\rangle = \begin{bmatrix} 1 & 0 \end{bmatrix}$ and $\left|\mathbf{y}\right\rangle = \begin{bmatrix} 0 & 1 \end{bmatrix}$ denote the x and y linearly polarized modes, respectively. For other values of $\varphi_0$, the phase in Supplementary Equation (6) can be vectorized into elliptical polarizations accordingly.

It should be emphasized that there are two critical conditions for the realization of phase vectorization. One is that the light beam must possess inherently different polarization responses of the left and right circularly polarized modes; another is that the undesired circularly polarized modes of $\exp\left[i(2f(\theta,\varphi)-\beta)\right]\left|\mathbf{R}\right\rangle$ and $\exp\left[-i(2f(\theta,\varphi)-\beta)\right]\left|\mathbf{L}\right\rangle$ must be eliminated. When both the above conditions are satisfied, phase vectorization can be achieved, and the scalar phase in Supplementary Equation (6) can be linked directly with the vector polarization in Supplementary Equation (7).

## Part 2: Explanation of using the phase in Supplementary Equation 6

According to Supplementary Equations (4) and (5), the link between the polarization and the wavefront of a light beam is related to not only the phase information in Supplementary Equation (6) but also the beam amplitude, which can be expressed as

$$T = Amp\exp(i\phi), \tag{9}$$

Here, the parameter $\phi$ is shown in Supplementary Equation (6). The parameter *Amp* is the amplitude of the light beam, which can be expressed as

$$Amp=\text{Amplitude}[\cos\varphi_0\exp[i(f(\theta,\varphi)-\beta)+\sin\varphi_0\exp[-i(f(\theta,\varphi)-\beta)], \tag{10}$$

There are two reasons why only the phase in Supplementary Equation (6) is utilized in this paper. One is that the amplitude component in Supplementary Equation (9) can be neglected due to its small influence on the polarization created by the polarized-SLM. Here, we take an *m*=30-order VVB as an example to investigate the impact of the amplitude in Supplementary Equation (9). Supplementary Figure 2 presents the light intensities of the x-polarized beam created via phase and amplitude modulation, as shown in Supplementary Figure 2(a), and only phase modulation, as shown in Supplementary Figure 2(h), where Supplementary Figure 2(a) illustrates the accurate link in Supplementary Equation (9) and Supplementary Figure 2(h) is obtained from the approximate link in Supplementary Equation (6). By comparing the x, y component and total light intensities after the pinhole in Supplementary Figures 2(b, c, d) with that of Supplementary Figures 2(i, j, k), the accurate and approximate links are almost identical, with a deviation of only 0.01%. That is, the polarizations in Supplementary Figures 2(e, f, and g) are consistent with those in Supplementary Figures 2(l, m, and n). Thus, the amplitude component in Supplementary Equation (9) can be neglected without affecting the validity of the polarized SLM.

The other reason is that the amplitude component in Supplementary Equation (9) affects the uniformity of the output light beam. Supplementary Figure 3 shows the light beams with linear polarization in the background and circular polarizations in the pixel-like zone, which are created by amplitude and phase modulation (Supplementary Figure 3(a)) and phase modulation (Supplementary Figure 3(e)), respectively. Due to the amplitude modulation in Supplementary Equation (9), the light intensity of linear polarization in the background is less than that of circular polarization by a factor of 2, as shown in Supplementary Figure 3(b). By contrast, the light beam created by the phase shown in Supplementary Figure 3(e) can not only obtain polarizations as those in Supplementary Figure 3(b, c, and d) but also maintain the uniformity of the whole light beam, as shown in Supplementary Figures 3(f, g, and h). Normally, a polarized-SLM is required to adjust the polarization without affecting the light intensity. Thus, for the above two reasons, only the phase in Supplementary Equation (6) is utilized in this paper.

## Supplementary Note 2: Theoretical principle of polarized-SLM

In this note, we present the entire theoretical principle of polarized-SLM and all the theoretical results of the experiments in the main text. As shown in Supplementary Figure 4, the polarized-SLM in Figure 2 can be simplified into a single filter system composed of a focusing system and a reconstruction system. Here, PH denotes a pinhole, and $OL_1$ and $OL_2$ are two objective lenses.

## Part 1: Focusing system

In the focusing system, the incident $m$=30-order VVB is composed of a right and a left circularly polarized mode with an inverse vortex [see Equation (1)]. When modulated by the phase $\phi = \pm(m\varphi - \beta)$, VVB exhibits different polarization responses and divides into desired and undesired polarized modes [See Equations (2) and (3)]. Here, we demonstrate the spatial separation of desired and undesired polarized modes in the focal region of $OL_1$.

Based on the Debye vectorial diffraction theory, the electrical fields near the focal point can be expressed as [6]

$$\mathbf{E}_f = \mathcal{F}\left\{P(\rho)T_p l_0(\theta)\cos^{1/2}\theta \mathbf{M}\mathbf{E}_i \exp(-ik_z z)/\cos\theta\right\}, \tag{11}$$

where $\theta$ is the convergent angle, and the maximum convergent angle $\alpha = \arcsin(NA/n)$. $NA$ is the numerical aperture of $OL_1$, and $n$ is the refractive index in the focusing space. $k_z = k\cos\theta$ is the $z$ component of the wavevector $k = 2n\pi/\lambda$, where $\lambda$ is the wavelength of the incident VVB. $\mathcal{F}$ denotes the Fourier transform, and $P(\rho)$ represents the $OL_1$ aperture, which can be expressed as follows:

$$P(\rho) = \begin{cases} 1 & 0 < \rho < R \\ 0 & otherwise \end{cases}, \tag{12}$$

where $R$ is the radius of $OL_1$. $T_p = \exp(i\phi)$, and the phase $\phi$ can be found in Equation (6). $l_0(\theta)$ denotes the electrical amplitude of the incident VVB. Due to the slight deviation between the conjugated plane of the phase-only SLM and VP, the output vector beam from the polarized-SLM has a hollow shape. In this case, $l_0(\theta)$ can be approximately expressed as follows:

$$l_0(\theta) = \begin{cases} 0 & 0 < \sin\theta/NA < 0.2 \\ 1 & 0.2 \le \sin\theta/NA < 1 \end{cases}. \tag{13}$$

In Supplementary Equation (11), $\mathbf{E}_i$ represents the polarization state of the VVB. For the sake of simplicity, the $m$=30-order VVB in Equation (1) is simplified into a superposition of the x and y polarized modes, expressed as follows:

$$\mathbf{E}_i = \cos(m\varphi)|\mathbf{x}\rangle + \sin(m\varphi)|\mathbf{y}\rangle, \tag{14}$$

where $|\mathbf{x}\rangle$ and $|\mathbf{y}\rangle$ denote the x and y linearly polarized modes, respectively, and $\mathbf{M}$ denotes a polarization transformation matrix caused by OL$_1$, which can be expressed as [6]

$$\mathbf{M} = \begin{bmatrix} \cos^2\varphi\cos\theta + \sin^2\varphi & (\cos\theta-1)\sin\varphi\cos\varphi & -\sin\theta\cos\varphi \\ (\cos\theta-1)\sin\varphi\cos\varphi & \cos^2\varphi + \sin^2\varphi\cos\theta & -\sin\theta\sin\varphi \\ \sin\theta\cos\varphi & \sin\theta\sin\varphi & \cos\theta \end{bmatrix}. \tag{15}$$

Eventually, the focal light intensity of the VVB can be obtained using $I = \left|\mathbf{E}_f\right|^2$.

**Spatial separation of desired and undesired polarized modes**

In the following simulations, $NA$=0.01, and $n$=1. The unit of length in all the figures is the wavelength $\lambda$, and the light intensity is normalized to the unit value. Because the absolute values of the topological charges in Equations (2) and (3) are equal for the desired and undesired polarized modes, the positions of the polarized modes in Equation (2) are identical to those of their counterparts in Equation (3). Here, we take only the light beam in Equation (2) as an example to investigate the spatial separation of the desired and undesired polarized modes in the focal region of OL$_1$.

Supplementary Figure 5 represents the focal light intensity of the VVB modulated by the phase $\phi = m\varphi - \beta$ with $m$=30 and (b) $\beta = 0$;(c) $\beta = 5\varphi$. That is, the topological charges of the left and right circularly polarized modes in Equation (2) are 0 and 60 for Supplementary Figures 5(b) and 5, respectively, and 55 for Supplementary Figure 5(c). Normally, both the left and right polarized modes are entangled during propagation in free space. Consequently, the phase-to-polarization link has been thought to be impossible. However, when focusing on OL$_1$, different topological charges lead to different positions in the focal region, thereby permitting the spatial separation of the left and right polarized modes [7]. If only the order of VVB is large enough, the left and right polarized modes can be spatially separated. Specifically, the desired left circularly polarized mode is located at the center, while the undesired right circularly polarized mode is at the outer ring position, as shown in Supplementary Figure 5 (b and c). Thus, one can retain the desired polarized mode simply by eliminating the undesired mode using a pinhole, which is the key to phase vectorization.

Note that the pinhole must be placed between the desired and undesired polarized modes so that the undesired mode is eliminated without affecting the desired mode. Therefore, the active area of the desired polarized mode is determined by the size of the pinhole. Generally, a larger active area is better. In Supplementary Figure 5, the VVB modulated by the phase $\phi = \pm(m\varphi - \beta)$ with $\beta = \sigma\varphi$ can be expressed as

$$\mathbf{E}_l = \exp\left[i(2m - \sigma)\varphi\right]|\mathbf{R}\rangle + \exp(-i\sigma\varphi)|\mathbf{L}\rangle, \tag{16}$$

$$\mathbf{E}_r = \exp\left[-i(2m - \sigma)\varphi\right]|\mathbf{L}\rangle + \exp(i\sigma\varphi)|\mathbf{R}\rangle, \tag{17}$$

where $|\mathbf{L}\rangle = \begin{bmatrix} 1 & i \end{bmatrix}$ and $|\mathbf{R}\rangle = \begin{bmatrix} 1 & -i \end{bmatrix}$ denote the left and right circularly polarized modes, respectively. As shown in Supplementary Figure 5(a), the larger the difference in topological charge $\Delta l = |2m - \sigma| - |\sigma|$, the greater the distance between the inner and outer polarized modes. A larger distance leads to a larger pinhole size, thereby increasing the active area of the desired polarized mode. There are two methods for enlarging the size of the pinhole.

**First method**

When $\sigma > 0$, $\Delta l$ becomes smaller as $\sigma$ is incremented, thus, only a relatively small pinhole size can be obtained. However, when $\sigma \leq 0$, no matter how large $|\sigma|$ is, one can always obtain an invariant $\Delta l = |2m|$. As $|\sigma|$ increases, the topological charges of the outer and inner polarized modes increase accordingly. For this reason, the pinhole must be enlarged to avoid affecting the desired polarized mode. That is, a large active area of the desired polarized mode can be achieved. It should be emphasized that the above case is only valid when $\sigma \leq 0$. Suppose that the invariant $\Delta l = |2m|$ could also be achieved when $\sigma > 0$ and that the active area of the desired polarized mode is no longer restricted by the pinhole. According to Supplementary Equations (16) and (17), the inner desired polarized modes are $\exp(-i|\sigma|\varphi)|\mathbf{L}\rangle$ and $\exp(i|\sigma|\varphi)|\mathbf{R}\rangle$ when $\sigma > 0$, while the modes when $\sigma \leq 0$ are $\exp(i|\sigma|\varphi)|\mathbf{L}\rangle$ and $\exp(-i|\sigma|\varphi)|\mathbf{R}\rangle$. Thus, the key to achieving an identical $\Delta l = |2m|$ when $\sigma > 0$ lies in how to convert the desired inner polarized modes $\exp(i|\sigma|\varphi)|\mathbf{L}\rangle$ and $\exp(-i|\sigma|\varphi)|\mathbf{R}\rangle$ when $\sigma \leq 0$ into $\exp(-i|\sigma|\varphi)|\mathbf{L}\rangle$ and $\exp(i|\sigma|\varphi)|\mathbf{R}\rangle$ when $\sigma > 0$.

This polarization conversion can easily be achieved with the aid of one half-waveplate (HWP). When passing through the HWP, the desired polarized modes $\exp(i|\sigma|\varphi)|\mathbf{L}\rangle$ and $\exp(-i|\sigma|\varphi)|\mathbf{R}\rangle$ for the case of $\sigma \leq 0$ can be expressed as follows:

$$\mathbf{E}_{outL} = \exp\left(i|\sigma|\varphi\right)\mathbf{H}|\mathbf{L}\rangle, \tag{18}$$

$$\mathbf{E}_{outR} = \exp\left(-i|\sigma|\varphi\right)\mathbf{H}|\mathbf{R}\rangle, \tag{19}$$

where $\mathbf{H}$ is the Jones matrix of HWP and can be written as

$$\mathbf{H} = \begin{bmatrix} \cos 2\psi & \sin 2\psi \\ \sin 2\psi & -\cos 2\psi \end{bmatrix}. \tag{20}$$

Here, $\psi=0$ represents the angle between the fast axis and the x-axis. The output polarized modes in Supplementary Equations (18) and (19) can further be simplified to

$$\mathbf{E}_{outL} = \exp\left(i|\sigma|\varphi\right)|\mathbf{R}\rangle, \tag{21}$$

$$\mathbf{E}_{outR} = \exp\left(-i|\sigma|\varphi\right)|\mathbf{L}\rangle. \tag{22}$$

Supplementary Equations (21) and (22) indicate that the desired polarized modes of the VVB modulated by $\phi = \mp(m-\sigma)\varphi$ when $\sigma \le 0$ can be converted into that of the VVB modulated by $\phi = \pm(m-\sigma)\varphi$ when $\sigma > 0$ with the aid of HWP. Because the invariant $\Delta l = |2m|$ can be obtained by the above polarization conversion, a large active area can be achieved for the desired polarized mode.

**Second method**

The second method, which is more direct, is to increase the $m$-order of the VVB. For example, when $m=30$ and $\sigma=10$, the desired and undesired polarized modes are 136.22 μm and 563.24 μm away from the geometric focus of OL₁, respectively. Thus, the radius of the pinhole can be adjusted to 400 μm. When $m=60$ and $\sigma=10$, the desired and undesired polarized modes are 136.22 μm and 1186.88 μm away from the geometric focus of OL₁, respectively, and the pinhole radius can be increased up to 800 μm. A larger $m$ leads to a larger pinhole size. Thanks to the mature Q-plate technique [8-10], the $m$-order of the VVB can reach as high as 126. Thus, the desired and undesired polarized modes are 136.22 μm and 2537.06 μm away from the geometric focus of OL₁, and the pinhole can be adjusted to the inconceivable radius of 2000 μm. Combining the above two methods, one can always obtain sufficient room for the polarization and phase modulation of polarized-SLM.

# Part 2: Reconstruction system

Part 1 demonstrated that $m$-order VVB modulated by the phase in Equation (6) is spatially separated into left and right circularly polarized modes located at different positions in the focal region of OL₁ [See Supplementary Figure 5]. Thus, the desired polarized mode can be extracted by filtering out the undesired mode using a pinhole. After passing through the objective lens OL₂, the desired polarized mode is reconstructed, and its electric field can be expressed as [6]

$$\mathbf{E} = \left( P(\rho) \cos^{1/2} \theta \right)^{-1} \mathbf{M}^{-1} \exp(ik_z z) \cos\theta \mathcal{F}^{-1} \left( \mathbf{E}_p \right), \tag{23}$$

where $\mathcal{F}^{-1}$ denotes the inverse Fourier transform, and $\mathbf{E}_p$ is the electric field behind the pinhole in Supplementary Figure 4, namely, the desired polarized mode. $\mathbf{M}^{-1}$ is the inverse polarization transformation matrix of the reconstructive lens OL₂, which can be expressed as [6]

$$\mathbf{M}^{-1} = \begin{bmatrix} \cos^2\varphi\cos\theta + \sin^2\varphi & (\cos\theta - 1)\sin\varphi\cos\varphi & \sin\theta\cos\varphi \\ (\cos\theta - 1)\sin\varphi\cos\varphi & \cos^2\varphi + \sin^2\varphi\cos\theta & \sin\theta\sin\varphi \\ -\sin\theta\cos\varphi & -\sin\theta\sin\varphi & \cos\theta \end{bmatrix}. \tag{24}$$

Finally, the light intensity output from OL₂ can be obtained using $I = |\mathbf{E}|^2$.

Generally, there are three types of polarization: left and right circular polarization, and linear polarization. Note that elliptical polarization is a combination of circular and linear polarization. Supplementary Figure 6 presents the theoretical results of Figure 3, which demonstrates that these three polarizations can link with three particular phases using phase vectorization. Specifically, the forward and reverse vortex phases can be vectorized into left and right circular polarizations, respectively, while the binary phase corresponds to linear polarization whose direction can be adjusted by the relative displacement of the phase structure.

To explain the principle of phase vectorization more clearly, we calculate the light intensities behind the pinhole $I = |\mathbf{E}_p|^2$ and their corresponding reconstruction polarizations from OL₂ in Supplementary Figures 7. The theoretical predictions in Supplementary Figure 7 are consistent with the experimental results in Supplementary Figure 8. Taking the reconstruction polarization in Supplementary Figure 7 (b) as an example, as demonstrated in Part 1, the undesired polarized mode is located in the outer ring, which is far from the desired inner ring. Thus, the undesired polarized mode is eliminated by the pinhole, and only the desired polarized mode is retained (see Supplementary Figure 7 (m). Because the desired polarized mode is the result of focusing the VVB with the phase in Supplementary Figure 7 (a), the desired polarized mode has a one-to-one correspondence with the VVB phase, which further establishes a direct link between the phase and the reconstruction polarization as shown in Supplementary Figure 7 (b). In this way, the VVB phase can be vectorized into arbitrary polarization.

As shown in Supplementary Figure 4, the VVB phase can be pixelated by phase-only SLM. Pixelate polarization can be achieved dynamically and in real time by the polarized-SLM. To demonstrate this, we present the theoretical results of Figure 5 in Supplementary Figure 9, which has four pixel-like zones named A, B, C, and D in the light beam, whose sizes are the same as those in Supplementary Figure 7. Supplementary Figure 9 (a, d, and g) show the corresponding VVB phases. In these four zones, one can even create four different VVB orders [see Equation (8)], verifying that pixelate phase manipulation of the VVB enables pixelate polarization adjustment of the light beam. In Supplementary Figure 9, subfigures (j, k, and l) show the light intensities behind the pinhole, which are utilized to reconstruct the polarizations in subfigures 9 (b, e, and h).

# Supplementary Note 3: Arbitrary vector light beam

This note presents three examples of vector light beams created using polarized-SLM. According to Equation (6), the phase for linear polarization can be simplified to

$$\phi = \text{Phase}\left\{\cos[(m\varphi - \beta)]\right\}, \tag{25}$$

where $m = 30$ is the order of the VVB; $\beta$ denotes the angle between the polarization direction and the x axis. The corresponding linear polarization can be written as follows:

$$\mathbf{E} = \cos\beta |\mathbf{x}\rangle + \sin\beta |\mathbf{y}\rangle, \tag{26}$$

where $|\mathbf{x}\rangle = [1 \quad 0]$ and $|\mathbf{y}\rangle = [0 \quad 1]$ denote the x and y linearly polarized modes, respectively. Because the parameter $\beta$ can be adjusted at will using the phase-only SLM in Figure 2, the phase in Supplementary Equation (25) can be vectorized into a linearly polarized beam with a designed polarization distribution.

## Case 1: Arbitrary order VVB

In this case, the parameter $\beta = n\varphi$ and the polarization in each pixel is adjusted to the direction of $\beta = n\varphi$. Thus, the phases in Supplementary Figure 10 (a, d, and h) can be vectorized into $n$-order VVBs, which can be expressed as

$$\mathbf{E} = \cos n\varphi |\mathbf{x}\rangle + \sin n\varphi |\mathbf{y}\rangle, \tag{27}$$

where $n=1$, 2, and 3 in Supplementary Figure 10 (b, e, and i), respectively. Supplementary Figure 10 (c, f, and j) are the corresponding light intensities passing through a polarizer (purple arrows).

## Case 2: Vector beam with multiple zones

In this case, the parameter $\beta$ is divided into three zones, namely, zones A, B, and C, as shown in Supplementary Figure 11 (a and e). The polarization in each zone can be manipulated individually using polarized-SLM.

Supplementary Figure 11 (b) presents the experimental result of a light beam created by the phase in Supplementary Figure 11 (a), whose polarization state can be expressed as follows:

$$\mathbf{E} = \begin{cases} \mathbf{VVB}_{n=1} & Azone: \quad 0 < r \leq R/3 \\ \mathbf{VVB}_{n=2} & Bzone: R/3 < r \leq 2R/3 \\ \mathbf{VVB}_{n=3} & Czone: \quad 2R/3 < r \leq R \end{cases}, \tag{28}$$

where $\mathbf{VVB}_n$ denotes VVB with the order $n=1$, 2, and 3 for zones A, B, and C, respectively.

Supplementary Figure 11(c, d) shows the light intensities of the A, B, and C zones passing through the polarizer (indicated by the purple arrows).

Likewise, Supplementary Figure 11(f) presents the experimental result of the light beam created by the phase in Supplementary Figure 11(e), the polarization state of which can be expressed as

$$\mathbf{E} = \begin{cases} \mathbf{LP}_{\beta=0} & Azone: \quad 0 < r \leq R/3 \\ \mathbf{LP}_{\beta=-0.25\pi} & Bzone: R/3 < r \leq 2R/3, \\ \mathbf{LP}_{\beta=-0.5\pi} & Czone: \quad 2R/3 < r \leq R \end{cases} \tag{29}$$

where $\mathbf{LP}_\beta$ denotes linearly polarized light whose polarization direction is $\beta$. Supplementary Figure 11(g, h) shows the light intensities of the A, B, and C zones passing through the polarizer (indicated by the purple arrows).

## Case 3: Special vector beam

In this case, the phases in Supplementary Figure 12(a, b, c, and d) are vectorized into the special vector beams in Supplementary Figure 12(e, f, g, and h), respectively, whose polarized state can be expressed as

$$\mathbf{E} = \cos \beta |\mathbf{x}\rangle + \sin \beta |\mathbf{y}\rangle, \tag{30}$$

where the parameter $\beta = 2\eta\pi \sin\theta / NA + \delta\varphi$ with $\eta = 2$, $NA = 0.01$ and (a) $\delta = 0$; (b) $\delta = -1$; (c) $\delta = -2$; and (d) $\delta = -3$, respectively. Supplementary Figure 12(i, j, k, and l) shows the light intensities of a special vector beam passing through the polarizer (indicated by the purple arrows).

# Supplementary Note 4: Additional discussion of polarized-SLM based on phase vectorization

## Part 1: Technical discussion of polarized-SLM

As the core of polarized-SLM in Figure 2, phase vectorization is capable of vectorizing the scalar phase into vector polarization. In this way, pixel-level polarization modulation can be achieved dynamically and in real time. Suppose an additional phase $\psi$ is superposed with the phase $\phi$ in Equation (6). Then, the entire VVB phase can be expressed as $\Omega = \phi + \psi$. According to the essence of phase vectorization, the first term, $\phi$, represents three particular phases: binary phase, vortex phases, and combinations of both, which correspond to linear, circular, and elliptical polarization, respectively. The second term $\psi$ represents the pure phase modulation of the light beam. After the polarization state is determined by the first term $\phi$, the light beam is modulated by the phase $\psi$. Therefore, polarized-SLM can not only fully manipulate the polarization but also retain complete phase adjustment of the light beam.

Although polarized-SLM simultaneously enables both polarization and phase modulation, the active area is determined by the pinhole. On the premise of eliminating the undesired polarized mode, the pinhole must be sufficiently large to accommodate the phase and polarization modulation of the desired polarized mode. There are two methods for increasing the size of the pinhole [See Supplementary Note 2]. One is to enlarge the $m$-order of the VVB. Benefitting from the mature technique of Q-plate [8-10], the $m$-order of the VVB can reach as high as 126. In this case, the inner and outer polarized modes are 136.22 μm and 2537.06 μm away from the geometric focus of $OL_1$. Thus, the pinhole can be set to an inconceivable radius of 2000 μm. Another method is based on the conversion of the polarized mode with the aid of a half-waveplate (HWP). In this case, the topological charge difference between the undesired and desired polarized modes remains the same, namely, $\Delta l = |2m|$. By combining both methods, sufficient room can always be obtained for the polarization and phase modulation of polarized-SLM.

Energy efficiency is another factor affected by the pinhole. When modulating by the phase in Equation (6), the $m$-order VVB is divided into two parts: the desired and undesired polarized modes. Both polarized modes have equal energy. During the phase vectorization process, the undesired polarized mode is always eliminated by the pinhole while the desired mode is retained. Thus, the overall energy efficiency of polarized-SLM is 50%, which is almost ten times as high as that of the methods based on the interference principle (which normally achieve an efficiency of only a few percent) [11, 12].

## Part 2: Difference between phase vectorization and conventional polarization modulation principle

Polarization manipulation is an important topic in classical optics. There are many works that have reported on how to manipulate the polarization of light beams. Here, we only note some methods that truly create arbitrary polarizations in free space [11-22]. For example, an interferometric optical system is a common scheme in which two orthogonally polarized beams with different phases are coherently superposed to form a light beam with arbitrary polarization [11-16]. The polarization exiting a birefringent waveplate can be adjusted when the phase accumulated by the light polarized parallel to the extraordinary or ordinary axis of a waveplate is changed. Similarly, a metasurface with effective birefringence can achieve full polarization adjustment by controlling the phases of two electric components $\mathbf{E}_x$ and $\mathbf{E}_y$ [21, 22]. Phase-only SLMs also exhibit a similar phenomenon. One can only operate on one polarization component at a time and delay the components in phase, thereby creating a light beam with a designed polarization [17-20].

Supplementary Figure 13 presents the difference between the polarized-SLM based on phase vectorization, a waveplate with a birefringent effect and an interferometric optical system. As shown in Supplementary Figure 13(b), after passing through a waveplate, the incident light beam is split into two orthogonally polarized light beams, namely, extraordinary and ordinary light beams. Here, we call extraordinary and ordinary light beams $o$ light and $e$ light, respectively. The phase of $e$ light can be adjusted by changing the thickness of the waveplate. Because both $o$ light and $e$ light propagate coaxially in the waveplate, the polarization output from the waveplate is the result of coherently superposing $o$ light and $e$ light. This kind of coaxial interferometric scheme can also be realized by using a metasurface with effective birefringence [21, 22] or using a phase-SLM [17-20]. Theoretically, polarizations generated by the above methods are, in essence, the same as those of the common interferometric optical system [11-16] as a result of superposing two orthogonally polarized beams, as shown in Supplementary Figure 13(c). It should be emphasized that the phase that affects the polarization in the above methods is entirely different from that of a polarized-SLM based on phase vectorization. Taking the waveplate as an example, the phases of $o$ light or $e$ light cannot change the polarization of $o$ light and $e$ light, respectively. The polarization is only the result of superposing two independent orthogonally polarized beams with different phases. Thus, these above methods are very simple and typically involve two steps: one is the generation of two orthogonally polarized beams, and the other is the adjustment of the phases of both light beams followed by the superposition of the beams

By contrast, phase vectorization, as depicted in Supplementary Note 1, has a one-to-one correspondence between the phase and polarization of a single light beam. As shown in Supplementary Figure 13(a), phase vectorization is actually a process of inherent polarized mode extraction for light beam. In classical optics, every vector beam is composed of two pairs of inherent polarized modes, namely, the left and right circularly polarized modes or x- and y-linear polarized modes. For example, $m$-order VVB can be considered a combination of the left and right circularly polarized modes from Equation (1) or the x- and y-linear polarized modes from Supplementary Equation (14). The left and right circularly polarized modes form a linearly polarized beam in free space. Normally, both inherent polarized modes are entangled when propagating in free space. Thus, extracting only one of the polarized modes from a light beam has always been considered impossible in physics. For example, we cannot obtain the left circularly polarized mode from a linearly polarized beam because both the left and right circularly polarized modes cannot be separated in free space. From this perspective, although a polarized-SLM based on phase vectorization is achieved by simply using a phase-only SLM and a Q-plate, the physical problem solved in this work is fundamentally different from those considered in the above methods based on the superposition of two independent polarized beams. Note that the Q-plate is merely an optical element for the realization of a suitable vector beam.

Through the extraction of the polarized mode of a light beam, phase vectorization, namely, the phase-to-polarization link, can vectorize three specific phases—the binary phase, the vortex phase, and the combination of both phases—into linear, circular and elliptical polarizations, respectively. Unlike all the present methods, the polarized-SLM approach based on this new principle of polarization modulation does not require a complicated algorithm, highly interferometrically precise alignment, a complex optical system or an expensive and difficult fabrication process; additionally, this method promotes the dynamic and real-time adjustment of the polarization of light beams. To the best of our knowledge, these properties cannot be achieved by optical devices based on the birefringent effect [21, 22].

**Part 3: The last step in unifying the amplitude, phase and polarization**

Heretofore, there are three relationships among the amplitude, phase and polarization in classical optics, namely, the mutual link between the phase and amplitude, the polarization-to-amplitude link and the polarization-to-phase link. As depicted in Supplementary Figure 1, the mutual link between the phase and amplitude can be realized by wavefront modulation techniques and has been the cornerstone of scalar optics in the past century. Malus's law represents the polarization-to-amplitude link, which transforms polarization modulation into amplitude adjustment using a polarizer. The third optical link demonstrated by the PB phase is the polarization-to-phase link. These three links imply that vector polarization can link to the scalar phase and amplitude, but the opposite is not true. That is, these three properties of a light beam are relatively independent. Phase vectorization based on the extraction of the inherent polarized mode can establish a previously unimaginable phase-to-polarization link by transforming all the polarization states into three specific phases. By combining the above three relationships with the principle of phase vectorization, the phase, amplitude and polarization of a light beam can be adjusted simultaneously merely by the phase. Thus, the phase-to-polarization link can be considered the last step in unifying the amplitude, phase and polarization.

From the above discussions, establishing the phase-to-polarization link is a fundamental physical breakthrough that provides a new principle for polarization modulation, and the results can be used to promote extensive developments in optics and pave the way for the era of vector optics.

## Supplementary References